\documentstyle[12pt]{article}
\begin{document}
\begin{titlepage}
  \begin{flushright}
    KUNS-1658\\[-1mm]
    YITP-00-18\\[-1mm]
    hep-ph/0004175\\[-1mm]
    April 2000
  \end{flushright}
  \vspace{5mm}
  
  \begin{center}
    {\large\bf New RG-invariants of Soft Supersymmetry Breaking
      Parameters}
    \vspace{1cm}
    
    Tatsuo~Kobayashi\footnote{E-mail address:
      kobayash@gauge.scphys.kyoto-u.ac.jp} and 
    Koichi~Yoshioka\footnote{E-mail address:
      yoshioka@yukawa.kyoto-u.ac.jp}
    \vspace{5mm}
    
    $^*${\it Department of Physics, Kyoto University
      Kyoto 606-8502, Japan}\\
    $^\dagger${\it Yukawa Institute for Theoretical Physics, Kyoto
      University,\\ Kyoto 606-8502, Japan}
    \vspace{1.5cm}
    
    \begin{abstract}
      We study new renormalization-group invariant quantities of soft
      supersymmetry breaking parameters other than the ratio of
      gaugino mass to gauge coupling squared by using the spurion
      method. The obtained invariants are useful to probe
      supersymmetry breaking and $\mu$-term generation mechanisms at
      high-energy scale. We also discuss the convergence behavior of
      fixed points of supersymmetry breaking parameters.
    \end{abstract}
  \end{center}
\end{titlepage}

Several types of supersymmetry (SUSY) breaking and mediation
mechanisms have been studied. Each mechanism leads to a proper
structure of SUSY breaking terms. The renormalization group (RG)
effects are not negligible to obtain SUSY breaking terms at low
energy. It is known that the ratio $M/g^2$, where $M$ is the gaugino
mass and $g$ is the gauge coupling, is RG-invariant at one-loop
level. For example, if $M/g^2$ for different gauge groups are measured
as similar values in future experiments, that implies the gaugino
masses are universal at the unification scale of gauge couplings. On
the other hand, when gauge interactions are relevant to the mediation
mechanism of SUSY breaking~\cite{DSB}, $M/g^2$ is given by the
group-theoretical factors for each gauge group and generally shows
disconnected structure with each other. Thus the RG-invariant quantity
is important to probe SUSY breaking mechanisms.

Recently, the renormalization group equations of soft SUSY breaking
parameters have been studied within the spurion
formalism~\cite{yamada}-\cite{kubo2}. Renormalization of softly broken
SUSY theories can be described by introducing the spurion fields into
rigid SUSY theories. That implies the beta-functions of soft SUSY
breaking parameters can be written by those of the gauge couplings and
anomalous dimensions of the rigid SUSY theories. This aspect has been
used to derive several important features and led to interesting
applications, e.g.\ calculations of higher-order beta functions of
soft SUSY breaking parameters, finiteness conditions, RG-invariant
trajectories and analytical solutions. As another example of
applications, in ref.~\cite{kubo2} the power-law behavior of running
soft SUSY breaking parameters has been derived within the framework of
extra dimensions.

In this letter, we consider a SUSY gauge-Yukawa system and derive new
RG-invariant quantities among soft SUSY breaking parameters through
RG-invariants among gauge and Yukawa couplings by use of the spurion
technique. The RG invariance holds for any arbitrary value of
couplings, not only on a specific RG trajectory. Such invariants of
soft SUSY breaking parameters are useful to probe the SUSY breaking
mechanisms at high energy. Furthermore we discuss the convergence of
infrared fixed points from the obtained RG invariants.

Let us consider the $N=1$ SUSY gauge theory with the gauge coupling
$g$ and the following superpotential,
\begin{equation}
  W \;=\; y\, \Phi_1 \Phi_2 \Phi_3.
\end{equation}
The one-loop beta functions of the gauge and the Yukawa couplings are
obtained
\begin{eqnarray}
  {d\alpha\over dt} &=& b\alpha^2,
  \label{b-g}\\
  {d\alpha_y\over dt} &=& \alpha_y \bigl(a\alpha_y -c\alpha\bigr),
  \label{b-y}
\end{eqnarray}
where $\alpha\equiv g^2/(16\pi^2)$ 
and $\alpha_y\equiv y^2/(16\pi^2)$. The coefficients, $a$, $b$ and $c$
are written in terms of group-theoretical factors and $a$ and $c$ are
always positive. The asymptotically free (non-free) theory corresponds
to $b<0$ ($b>0$).

The soft SUSY breaking terms are written
\begin{equation}
  -{\cal L}_{\rm soft} \;=\; \left[{M\over 2}\lambda\lambda +Ay \phi_1
    \phi_2 \phi_3 + {\rm h.c.} \right] +\sum_i m_i^2 |\phi_i|^2,
  \label{soft-L}
\end{equation}
where $\lambda$ is the gaugino and $\phi_i$ is the scalar field in the
chiral multiplet $\Phi_i$. The total Lagrangian with the soft SUSY
breaking terms (\ref{soft-L}) can be written in terms of $N=1$
superfields by introducing the spurion fields $\eta=\theta^2$ 
and $\bar\eta=\bar\theta^2$~\cite{spurion}. Furthermore, the beta
functions of $M$, $A$ and $\Sigma$, where
\begin{equation}
  \Sigma \;\equiv\; m_1^2+m_2^2+m_3^2,
\end{equation}
are obtained through those of $\alpha$ and $\alpha_y$. That is, in
eqs.~(\ref{b-g}) and (\ref{b-y}) we replace $\alpha\to\tilde\alpha$
and $\alpha_y\to\tilde\alpha_y$ as follows~\cite{kazakov},
\begin{eqnarray}
  \tilde\alpha &=& \alpha(1+M\eta+\bar M\bar\eta
  +2M\bar M\eta\bar\eta),\\
  \tilde\alpha_y &=& \alpha_y (1-A\eta-\bar A\bar\eta 
  +(A\bar A+\Sigma)\,\eta\bar\eta),
  \label{exp-y}
\end{eqnarray}
and then we obtain the one-loop beta functions of SUSY breaking
parameters,
\begin{eqnarray}
  {dM\over dt} &=& b\alpha M,
  \label{b-M}\\
  {dA\over dt} &=& a\alpha_y A+c\alpha M,\\
  {d\Sigma\over dt} &=& a\alpha_y (\Sigma+A\bar A)-2c\alpha M\bar M.
\end{eqnarray}
The two-loop and higher-loop beta functions are obtained similarly,
but those are scheme dependent~\cite{jjmvy}. Here we use the same
technique to derive RG-invariants among the SUSY breaking 
parameters $M$, $A$ and $\Sigma$.

Comparing the beta functions (\ref{b-g}) and (\ref{b-M}), we can
obtain the well-known RG-invariant, $M/\alpha$,
\begin{equation}
  {M\over\alpha} \;=\; \mbox{RG--invariant}.
  \label{RG-M}
\end{equation}
As an alternative derivation, we show this fact can be understood from
the spurion method. From eq.~(\ref{b-g}), we have the RG equation of
gauge coupling, $d\alpha^{-1}/dt\,=\,-b$ and then in the softly broken
case, the corresponding equation is
\begin{equation}
  {d\tilde\alpha^{-1}\over dt} \;=\; -b.
\end{equation}
Note that the right-hand side of the equation does not include any
couplings. The coefficient of $\eta$ in $\tilde\alpha^{-1}$ 
is $-M/\alpha$. Thus it is found that $M/\alpha$ is RG invariant. This
derivation implies that if we have a relation among supersymmetric
couplings, then we can extend it to the RG-invariants among soft SUSY
breaking parameters. As noted before, the RG-invariant $M/\alpha$ is
important for the purpose to relate values of $M$ between low and high
energies and to probe the SUSY breaking mechanism at high-energy
scale. Similarly, RG-invariants for other soft SUSY breaking
parameters, if they exist, are also helpful to probe SUSY breaking
mechanisms.

Now let us derive RG-invariants for $A$ and $\Sigma$ by use of the
spurion technique. The above illustration tells us that if we have a
relation of couplings like $f(\alpha,\alpha_y)=$ coupling-independent,
we can derive additional RG-invariants by expanding over the
Grassmannian variable $\eta$. First the comparison between the beta
functions of $\alpha$ and $\alpha_y$ (\ref{b-g},~\ref{b-y}) leads to
the RG-invariant $\Gamma$ of the rigid SUSY theory,
\begin{equation}
  \Gamma \;\equiv\; \left[1-{(b+c)\over a}{\alpha\over\alpha_y}
  \right]\alpha^{-(1+c/b)} \;\;=\; \mbox{RG--invariant}.
  \label{RGinv}
\end{equation}
Alternatively, we define a ratio between $\alpha$ and $\alpha_y$,
\begin{equation}
  R \;\equiv\; {a\over b+c}{\alpha_y\over\alpha},
\end{equation}
then the RG-invariant $\Gamma$ is written 
as $\Gamma\,=\,(R-1)/R\alpha^{1+c/b}$. The infrared fixed point of the
ratio corresponds to the point $R=1$. Now let us replace $\alpha$ 
and $\alpha_y$ in $\Gamma$ by $\tilde\alpha$ and $\tilde\alpha_y$ in
order to extend $\Gamma$ to the quantity $\tilde\Gamma$ including soft
SUSY breaking parameters,
\begin{equation}
  \tilde\Gamma \;=\; \Gamma-{X\over\alpha^{1+c/b}}\,\eta 
  -{\bar X\over\alpha^{1+c/b}}\,\bar\eta 
  +{Z\over\alpha^{1+c/b}}\,\eta\bar\eta \;\;=\; \mbox{RG--invariant},
\end{equation}
where
\begin{eqnarray}
  X &=& {A+M\over R}+\Bigl(1+{c\over b}\Bigr)\,{R-1\over R}M,\\
  Z &=& {\Sigma-M\bar M-(A+M)(\bar A+\bar M)\over R} 
  +\Bigl(1+{c\over b}\Bigr)\,{A+M\over R}\bar M \nonumber\\
  &&\qquad +\Bigl(1+{c\over b}\Bigr)\,{\bar A+\bar M\over R} M
  +\Bigl(1+{c\over b}\Bigr)\,{c\over b}\,{R-1\over R}M\bar M.
\end{eqnarray}
Thus we obtain the two more RG-invariants, $X/\alpha^{1+c/b}$ 
and $Z/\alpha^{1+c/b}$, among the soft SUSY breaking parameters. Note
that $X$ is a complex-valued and the RG-invariant $X/\alpha^{1+c/b}$
includes a $CP$ phase.

These new RG-invariants are important as well as $M/\alpha$ to probe
the SUSY breaking mechanism at high energy from experimental values if
soft SUSY breaking parameters would be measured in future. For
example, a certain type of SUSY breaking mechanism such as the
no-scale supergravity models~\cite{noscale} and the gauge-mediated
SUSY breaking models leads to suppressed values of $A$ at definite
energy scales (no-scale models also predict suppressions of $\Sigma$),
while string-inspired supergravity theories~\cite{string} lead 
to $A=-M$ and $\Sigma=M^2$ at high energy. The difference between
these initial conditions provides us with the meaningful difference
for the RG-invariants at low-energy scale,
$\Delta(X/\alpha^{1+c/b})=M_0/(R_0\alpha_0^{1+c/b})$ and 
$\Delta(Z/\alpha^{1+c/b}) \simeq |M_0|^2/(R_0\alpha_0^{1+c/b})$. Note
that since $X/\alpha^{1+c/b}$ and $Z/\alpha^{1+c/b}$ are RG invariant,
the low-energy differences can be written in terms of $\alpha_0$ and
$M_0$ given at the scale at which the boundary condition is imposed in
each case. For example, it is found from these that the difference is
larger for $b>0$ case by about 1 order of magnitude. Therefore one can
find the gauge-mediation scenarios could reveal rather different
signatures than others.

Furthermore the RG-invariants take certain values under a definite
condition. For example, on the RG-invariant trajectory we always have
$\Gamma=X=Z=0$, i.e.\ $R=1$. The quasi fixed-point solution (the large
Yukawa coupling case) corresponds to $R_0\gg 1$ and that leads 
to $X=(1+c/b)M_0$ and $Z=(1+c/b)c/b\,|M_0|^2$.

For another application, let us discuss the convergence behavior of
fixed points. As said above, the infrared fixed point of $R$ is
obtained as $R=1$. Similarly in the softly broken case, the fixed
points of $X$ and $Z$ correspond to the points $X=0$ and $Z=0$. That
implies
\begin{equation}
  A \;=\; -M,\qquad \Sigma \;=\; M\bar M,
\end{equation}
at the point $R=1$. From the beta functions of the gauge and Yukawa
couplings, we have
\begin{equation}
  {R-1\over R} \;=\; \xi\,{R_0-1\over R_0},
\end{equation}
with $\xi\equiv(\alpha_0/\alpha)^{1+c/b}$. The couplings with
subscript 0 denote some initial values for them. Hence it is found
that the convergence into $R=1$ is described by the 
quantity $\xi$~\cite{fpp}. In the asymptotically non-free gauge
theories, we have a rapid convergence to the fixed 
point $R=1$, i.e.\ a tiny value of $\xi$. Furthermore such analysis on
the convergence of the Yukawa fixed points has been extended to the
theories with extra spatial dimensions~\cite{extra-y}, where we can
have a good convergence even for asymptotically free theories with a
certain condition. As for the soft SUSY breaking parameters, we obtain
by replacing $R$ by $\tilde R$ (or from the RG invariance discussed
above),
\begin{equation}
  X \;=\; \xi X_0,\qquad Z \;=\; \xi Z_0.
\end{equation}
Remarkably, the convergence behaviors of $X=0$ and $Z=0$ are described 
by the same quantity $\xi$ as $R=1$. Therefore in the models where the 
Yukawa couplings go to the fixed points rapidly, the soft breaking
terms also have good convergence to their fixed points (at the same
rate).

So far we have considered the case that a single gauge coupling is
dominant in the anomalous dimensions of $\Phi$. In the minimal
supersymmetric standard model for example, the three gauge couplings
and the three gaugino masses contribute to RG equations, in particular
around the grand unification scale. For that purpose it is useful to
extend our analyses to the case with the gauge group $G=\prod G_i$
whose gauge couplings are $g_i$. The beta functions of the
supersymmetric couplings $\alpha_i=g_i^2/(16 \pi^2)$ and $\alpha_y$
are obtained
\begin{eqnarray}
  {d\alpha_i\over dt} &=& b_i\alpha_i^2,\\
  {d\alpha_y\over dt} &=& \alpha_y
  \Bigl(a\alpha_y-\sum_i c_i\alpha_i\Bigr).
\end{eqnarray}
In this case, the RG invariant quantity in the rigid theory is defined
by
\begin{equation}
  \Gamma \;\equiv\; {E\over\alpha_y}+aF \;=\; \mbox{RG--invariant},
\end{equation}
where $E\equiv\prod_i\alpha_i^{-c_i/b_i}$ 
and $F(t)=\int^t E(t')dt'$. Note that in the case of one gauge
coupling ($i=1$), this expression just reproduces the previous 
one (\ref{RGinv}) (multiplied by a constant). We extend it to the
softly broken RG-invariants by replacing $\alpha_i$ and $\alpha_y$
with the redefined couplings $\tilde\alpha_i$ and $\tilde\alpha_y$
where
\begin{equation}
  \tilde\alpha_i \;=\; \alpha_i(1+M_i\eta+\bar M_i\bar\eta 
  +2M_i\bar M_i\eta\bar\eta),
\end{equation}
and $\tilde\alpha_y$ is given in eq.~(\ref{exp-y}). From these, we can
find two RG-invariants $\Gamma_\eta$ and $\Gamma_{\eta\bar\eta}$,
which are the coefficients of $\eta$ and $\eta\bar\eta$ 
in $\tilde\Gamma$,
\begin{eqnarray}
  \Gamma_\eta &=& {E\over\alpha_y}
  \Bigl(A-\sum_i {c_i\over b_i}M_i\Bigr) -a\sum_i 
  \int^t E(t'){c_i\over b_i}M_i(t')dt',\\ 
  \Gamma_{\eta\bar\eta} &=& {E\over\alpha_y}
  \Bigl(A\bar A-\Sigma+E_{\eta\bar\eta}-A\sum_i {c_i\over b_i}\bar M_i 
  -\bar A\sum_i {c_i\over b_i}M_i\Bigr) 
  +a\int^t E(t')E_{\eta\bar\eta}(t')dt',
\end{eqnarray}
where $E_{\eta\bar\eta}$ is the coefficient of $\eta\bar\eta$ in $E$
and explicitly given by
\begin{equation}
  E_{\eta\bar\eta} \;=\; -\sum_i {c_i\over b_i}
  \Bigl(1-{c_i\over b_i}\Bigr)|M_i|^2 +\sum_{i>j}
  {c_i\over b_i}{c_j \over b_j}(M_i\bar M_j +M_j\bar M_i).  
\end{equation}

Finally, we consider the RG-invariants of the $\mu$ term and the
corresponding $B$ term for the application to the supersymmetric
standard models. The beta function of the $\mu$ term is given by
\begin{equation}
  {d\mu\over dt} \;=\; \mu\Bigl(a_\mu\alpha_y
  -\sum_i c_{\mu i}\alpha_i\Bigr).
\end{equation}
The coefficients $a_\mu$ and $c_{\mu i}$ are written in terms of the
group-theoretical factors and are always positive. The beta function
of the soft breaking parameter $B$ can be obtained by replacing the
supersymmetric couplings $\alpha_i$, $\alpha_y$ and $\mu$ 
by $\tilde\alpha_i$, $\tilde\alpha_y$ and $\tilde\mu$, where 
the $\eta$ coefficient of $\tilde\mu$ is given by $-B\mu$. In this
case we have the RG-invariant among the supersymmetric coupling,
\begin{equation}
  \mu E_\mu \alpha_y^{-a_\mu /a} \;\;=\; \mbox{RG--invariant},
\end{equation}
where
\begin{equation}
  E_\mu \;\equiv\; \prod_i \alpha_i^{-c'_i/b_i},
\end{equation}
with $c'_i = (a_\mu/a)\,c_i-c_{\mu i}$. By use of the spurion method,
this equation provides us with the RG-invariant concerned with $B$ as
\begin{equation}
  B+\sum_i {c'_i\over b_i}M_i -{a_\mu\over a}A \;=\;
  \mbox{RG--invariant}.
  \label{B}
\end{equation}
This simple RG-invariant would also be important to study the SUSY
breaking mechanisms. Moreover, since the $B$ parameter takes rather
different values depending on how the $\mu$ term is generated at
high-energy scale~\cite{mu}, we can use the RG-invariant (\ref{B}) for
selecting the $\mu$-term generation mechanisms. Note that the
invariant also includes a $CP$ phase as $X$.

To summarize, we have presented several RG-invariants of soft SUSY
breaking parameters, e.g.\ $X/\alpha^{1+c/b}$ and $Z/\alpha^{1+c/b}$
for the case with a single gauge coupling. The RG invariance holds not
only on a specific RG-trajectory but also for arbitrary set of
couplings. These are important to probe the SUSY breaking mechanisms
and the origin of $\mu$-term at high energy when the parameters would
be measured in future experiments. Our analyses have shown that if one
has a RG-invariant among SUSY couplings, one can generally derive
corresponding RG-invariants among soft SUSY breaking parameters. We
have also discussed the convergence behavior of the fixed points of
SUSY breaking parameters. The obtained results in this letter may be
extended to the cases with some Yukawa couplings~\cite{AM}.

\end{document}